\documentclass[acmsmall,screen]{acmart}
\AtBeginDocument{%
  }

\setcopyright{acmlicensed}
\usepackage{forest}
\usepackage{pgfplots}
\usepackage{makecell}
\usepackage{hyperref}
\copyrightyear{2018}
\acmYear{2018}
\acmDOI{XXXXXXX.XXXXXXX}
\acmConference[Conference acronym 'XX]{Make sure to enter the correct
  conference title from your rights confirmation email}{June 03--05,
  2018}{Woodstock, NY}
\acmISBN{978-1-4503-XXXX-X/2018/06}
\graphicspath{{./}}

\usepackage{booktabs}
\usepackage{tabularray}
\usepackage{tcolorbox}
\usepackage{enumitem}
\usepackage{xspace}
\usepackage{algorithm}
\usepackage{algpseudocode}
\usepackage{multirow}
\usepackage{pgfplots}
\usepackage{graphicx}
\pgfplotsset{compat=1.18}
\usepackage{tikz}
\usetikzlibrary{patterns}
\usepackage{tcolorbox}
\usepackage{xcolor}
\definecolor{box_bg_light}{HTML}{F1F8E9}    
\definecolor{frame_green}{HTML}{4CAF50}     

\newtcolorbox{mybox}{
  colback=box_bg_light,          
  colframe=frame_green,          
  width=\columnwidth,
  boxrule=0.3mm,
  top=2pt, bottom=2pt, left=3pt, right=3pt,
  before skip=6pt, after skip=6pt
}
\begin{document}

\title{\methodname: A Semantically-Rich and Efficient Online Sampler for Microservice Diagnostics}

\author{Yifan Yang}
\email{yifanyang6@link.cuhk.edu.cn}

\author{Aoyang Fang}
\email{aoyangfang@link.cuhk.edu.cn}

\author{Songhan Zhang}
\email{222010549@link.cuhk.edu.cn}

\author{Pinjia He}
\authornote{Corresponding author.}
\email{hepinjia@cuhk.edu.cn}
\affiliation{%
  \institution{The Chinese University of Hong Kong, Shenzhen}
  \city{Shenzhen}
  \country{China}
}

\renewcommand{\shortauthors}{Yang et al.}

\newcommand{\methodname}{Gleaner\xspace}

\begin{abstract}

Distributed tracing in microservices is critical for diagnostics but generates overwhelming data volumes, necessitating intelligent sampling. 
To maximize fidelity, state-of-the-art (SOTA) tail-based samplers analyze complete (or even log-enriched) traces by modeling them as graphs. 
However, this reliance on computationally expensive graph analysis creates a performance bottleneck that prohibits their use in online settings.

To this end, we propose \methodname, an online tail-sampling framework that breaks this trade-off. 
It is founded on the key insight that explicit graph structures are unnecessary for high-fidelity trace grouping. 
Instead, \methodname represents each trace as a "bag-of-edges" augmented with log semantics, replacing slow graph algorithms with highly efficient set-based operations. 
It also employs an alarm-driven quota and a diversity-preserving strategy to prioritize anomalous and rare traces for downstream Root Cause Analysis (RCA).
Experimentally, \methodname processes traces at 0.74ms each, improving Trace Pattern Coverage by up to 128.7\% and Shannon Entropy by up to 32.9\% over baselines. 
At just a 1\% sampling rate, \methodname improves RCA accuracy by 42\%-107\% over the next-best sampler. 
Moreover, RCA on \methodname's sampled data is more accurate than with the entire, unsampled dataset. 
This result reframes intelligent sampling from a data reduction technique to a powerful signal enhancement paradigm for automated operations.
\end{abstract}




\maketitle

\section{Introduction}

Microservice architectures, while enabling rapid development and deployment, introduce profound operational complexity~\cite{zhang2024failure,zhou-fault-2021,li-enjoy-2021,gan2021sage}. A single user request can trigger a cascade of interactions across hundreds of services, making system behavior difficult to comprehend~\cite{zhou2018fault}. Distributed tracing has emerged as an indispensable tool for observability, providing visibility into these complex request flows~\cite{yao2024chain,DeCaf2020,Beschastnikh2020,guo2020}. However, the sheer volume of data—often terabytes daily~\cite{dapper}—makes comprehensive storage and analysis infeasible. Furthermore, trace data typically follows a long-tail distribution, where a few common execution paths dominate. Consequently, intelligent \textbf{trace sampling} is not just a practical necessity but a critical component of modern observability pipelines.

The most straightforward approach, \textbf{head-based random sampling}, is widely considered inadequate. By making sampling decisions before the execution outcome is known, it disproportionately retains common, healthy traces while being highly likely to discard rare patterns~\cite{he2023steam}. Since system anomalies and critical failures often manifest as these rare traces, head-based methods risk discarding the very signals that engineers and SREs need most for diagnostics. This fundamental limitation has driven the community towards more sophisticated \textbf{tail-based sampling techniques}~\cite{las-casas-weighted-2018,las-casas-sifter-2019,huang-sieve-2021,he2023steam,huang2024trastrainer,xieshuaiyu-tracepicker-2025}. By waiting until a trace is complete, these methods can leverage holistic signals such as full call structure, end-to-end latency, and span status, and can bias the budget toward rare or abnormal traces, significantly improving the quality of the sampled data.

However, existing tail-based samplers suffer from a critical limitation: they rely almost exclusively on coarse, \textbf{span-level information}. They evaluate a trace based on its call structure (e.g., service A called B) and the final status of its constituent spans, overlooking the rich semantic details within each span's execution. This is a significant blind spot. The diagnostic value of a trace is often determined not by its high-level structure but by fine-grained events—such as error logs, retry attempts, or custom application events—that occur during a span's lifetime~\cite{zhang2022deeptralog,liu-uac-ad-2024,jiang2023look}. A sampler that is blind to this information might discard a trace that appears structurally normal (e.g., low latency, success status) but contains a critical error log, silently dropping the most vital signal for root cause analysis (RCA).

This disconnect is increasingly problematic as both industry and academia move towards multi-modal diagnostics. Open-source standards like OpenTelemetry are actively standardizing trace-log correlation and incorporating inner-span events into sampling strategies~\cite{opentelemetry-logging,opentelemetry-specificationoteps0265-event-visionmd}. Concurrently, advanced RCA algorithms increasingly leverage the semantic information in logs to improve diagnostic accuracy~\cite{yu2023nezha,gu2023trinityrcl,zhang2023robust,lee2023eadro,wang2021groot,sun2024art,wang2024mrca,sun-trace-log-clusterings-based-2023,wang-cross-system-2025,chen-root-2024,han-holistic-2024,zhang-etal-2024-mabc,han2024potential,xu2025openrca,fang_rethinking_2025}. Recognizing this, a handful of recent studies have attempted to incorporate logs into the sampling process~\cite{tian-itcrl-2024}. However, these pioneering methods rely on computationally expensive models like Heterogeneous Graph Neural Networks  (HGNNs), which require extensive training and are too slow for online, real-time processing in high-throughput environments~\cite{xieshuaiyu-tracepicker-2025,gao2025gnnsactuallyeffectivemultimodal}.

This reveals a critical \textit{fidelity-performance trade-off} in modern trace sampling: methods that are computationally efficient remain blind to semantically rich internal signals, while methods fine-grained enough to capture these signals are too slow for practical deployment. The central challenge, therefore, is to design a sampling mechanism that is both semantically rich and computationally tractable.

To bridge this gap, we propose \textbf{\methodname}, a novel online tail-sampling framework that breaks this trade-off. Our key insight is that explicit, computationally expensive graph structures are not necessary for high-fidelity, semantically-aware sampling. Instead, \methodname introduces a lightweight event-pair set (EPS) representation, modeling each trace as an efficient set of event pairs (\S\ref{sec:trace-encoder}). This representation captures both inter-span call dependencies and crucial intra-span events (e.g., logs, status codes) and replaces costly graph algorithms with highly efficient set operations. Building on this, \methodname incorporates an alarm-driven quota mechanism (\S\ref{sec:quota-allocator}) to prioritize resources during incidents and a diversity-preserving selector (\S\ref{sec:dpp-selector}) based on an optimized Determinantal Point Process (DPP) to ensure a rich and varied sample set. Compared to head-based sampling, \methodname provides vastly superior coverage of rare patterns. Compared to existing trace-only tail-samplers, it captures far richer semantic signals at a comparable or even faster speed. And critically, compared to emerging trace-and-log samplers, \methodname achieves its semantic richness without the prohibitive overhead of model training and inference, making it suitable for high-volume online deployment.

We conducted an extensive evaluation on a new, large-scale benchmark dataset.  Our results demonstrate that \methodname achieves high-throughput processing at 0.74ms per trace, while significantly outperforming state-of-the-art samplers on quality metrics. It achieves 11.6\%\textasciitilde{}128.7\% improvement in Trace Pattern Coverage and 2.8\%\textasciitilde{}32.9\% gains in Shannon Entropy over baselines. At a mere 1\% sampling rate, \methodname boosts the accuracy of downstream RCA tools by 42\%\textasciitilde{}107\% over the next-best sampler. Remarkably, RCA on \methodname's sampled data is more accurate than when using the entire, unsampled dataset. This finding reframes intelligent sampling from a mere data reduction technique to a powerful signal enhancement paradigm for modern automated operations.

The main contributions of this work are as follows:
\begin{itemize}
    \item We propose a novel method to construct a unified representation from traces and their associated logs, capturing call relationships, internal execution dynamics, and overall anomaly severity.
    \item We contribute a new large-scale dataset with 161 fault injection cases, over 1.4 million traces, and 517 distinct call paths, which will benefit future research in sampling and diagnostics.
    \item We present \textbf{\methodname}, the first practical and scalable online sampler that jointly leverages traces and logs, demonstrating its superior performance on an industrial-scale dataset.
    \item We show that a well-designed sampler not only accelerates downstream RCA tasks but can also significantly improve their accuracy, reframing sampling from a simple data reduction tool to an active signal enhancement strategy for AIOps pipelines.
\end{itemize}

\section{Background and Motivation}
\label{sec:background_and_motivation}

This section lays the groundwork for our research. We first establish why intelligent sampling is essential given the data characteristics of distributed tracing (§\ref{sec:necessity}). We then use concrete examples to expose the critical disconnect between semantically-blind samplers and the demands of modern diagnostic algorithms, which directly motivates the design of \methodname (§\ref{sec:disconnect}).

\subsection{The Challenge of Data Volume in Distributed Tracing}
\label{sec:necessity}

Modern microservice systems generate trace volumes reaching terabytes daily~\cite{dapper}, making comprehensive storage infeasible and necessitating intelligent sampling strategies. Critically, trace data exhibits a pronounced long-tail distribution over execution paths, where a \textit{path} is defined as a unique sequence of service operations invoked during request processing. In the Train-Ticket benchmark system~\cite{trainticket-paper}, the top 24.3\% of paths account for 82.3\% of total traces. This skew renders simple random sampling ineffective, as it would overwhelmingly retain redundant traces while discarding rare execution patterns. Consequently, intelligent, biased sampling strategies are essential~\cite{he2023steam, las-casas-sifter-2019, huang-sieve-2021, xieshuaiyu-tracepicker-2025}. Our work focuses on tail-based sampling, which makes decisions after trace completion, enabling strategies that prioritize valuable traces based on their characteristics~\cite{las-casas-sifter-2019, he2023steam, xieshuaiyu-tracepicker-2025}.

\subsection{The Semantic Gap in Trace Sampling}
\label{sec:disconnect}

A critical disconnect has emerged between trace sampling and diagnostics. While state-of-the-art samplers excel at prioritizing traces based on structural rarity or high latency, they remain semantically-blind; they operate on span-level metadata but ignore rich intra-span context like log events~\cite{huang-sieve-2021, xieshuaiyu-tracepicker-2025}. This creates a fundamental problem, as modern diagnostic algorithms are increasingly multi-modal, depending on the correlation between trace structures and log semantics for accurate root cause analysis~\cite{yu2023nezha, gu2023trinityrcl}.

\begin{figure}[t]
\centering
\resizebox{\columnwidth}{!}{%
\includegraphics{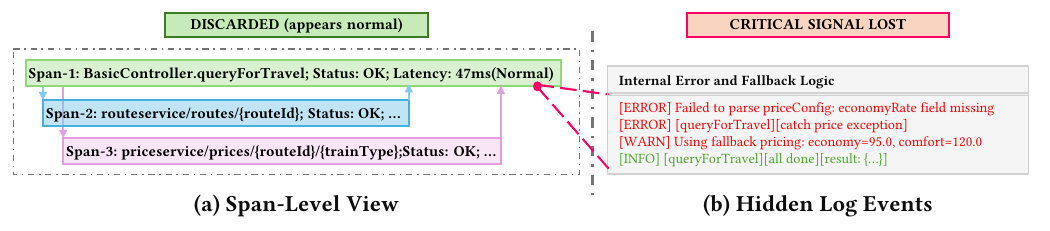}
}
\caption{Example trace illustrating the semantic blind spot of span-centric samplers. Despite normal span-level attributes (success status, acceptable latency), a critical ERROR log within the root span reveals a configuration parsing failure. Traditional samplers would discard this trace as routine, losing diagnostically valuable evidence.}
\label{fig:semantic_blind_spot}
\end{figure}

Figure~\ref{fig:semantic_blind_spot} provides a concrete example. 
A trace appears benign based on all span-level metrics (e.g., success status, normal latency) and would be discarded by conventional samplers. 
However, an embedded ERROR log reveals a critical configuration failure—precisely the kind of subtle issue that multi-modal tools are designed to detect. 
By discarding such traces, the sampler inadvertently filters out the crucial evidence that these downstream tools require, undermining the entire observability pipeline.

Table~\ref{tab:sota-comparison} systematically situates our work and highlights this gap. Existing methods have progressively incorporated more signals, yet only iTCRL and our proposed \methodname are aware of intra-span log events. Crucially, iTCRL's reliance on offline GNN training makes it unsuitable for real-time use. This analysis highlights the core technical challenge: designing a sampling mechanism that is both semantically rich and computationally tractable for high-throughput, online environments. \methodname is designed to fill this critical gap.

\begin{table}[t]
\centering
\caption{Comparison of Trace Sampling Methods and Their Capabilities.}
\label{tab:sota-comparison}
\resizebox{0.8\columnwidth}{!}{%
\begin{tabular}{@{}l|ccc|ccc@{}}
\toprule
\multirow{2}{*}{\textbf{Method}} & \multicolumn{3}{c|}{\textbf{Data Used}} & \multicolumn{3}{c}{\textbf{Key Capabilities}} \\
\cmidrule(lr){2-4} \cmidrule(lr){5-7}
& \textbf{Structure} & \textbf{Attributes} & \textbf{Extra} & \makecell[c]{\textbf{Anomaly} \\ \textbf{Aware}} & \makecell[c]{\textbf{Intra-span} \\ \textbf{Aware}} & \textbf{Online} \\
\midrule
\textbf{PEACH}~\cite{las-casas-weighted-2018} & \checkmark & & & & & \checkmark \\
\midrule
\textbf{Sifter}~\cite{las-casas-sifter-2019} & \checkmark & & & & & \checkmark \\
\midrule
\textbf{Sieve}~\cite{huang-sieve-2021} & \checkmark & \makecell[c]{Latency} & & \makecell[c]{Partial} & & \checkmark \\
\midrule
\textbf{TraceCRL}~\cite{TRACECRL} & \checkmark & \checkmark & & \checkmark & & \textcolor{ACMRed}{Offline training} \\
\midrule
\textbf{STEAM}~\cite{he2023steam} & \checkmark & \checkmark & & \checkmark & & \textcolor{ACMRed}{Offline training} \\
\midrule
\textbf{TraStrainer}~\cite{huang2024trastrainer} & \checkmark & \checkmark & \textcolor{ACMBlue}{Metrics} & \checkmark & & \checkmark \\
\midrule
\textbf{TracePicker}~\cite{xieshuaiyu-tracepicker-2025} & \checkmark & \checkmark & & \checkmark & & \checkmark \\
\midrule
\textbf{iTCRL}~\cite{tian-itcrl-2024} & \checkmark & \checkmark & \textcolor{ACMPurple}{Log events} & \checkmark & \checkmark & \textcolor{ACMRed}{Offline training} \\
\midrule
\textbf{\methodname (Ours)} & \checkmark & \checkmark & \textcolor{ACMPurple}{Log events} & \checkmark & \checkmark & \checkmark \\
\bottomrule
\end{tabular}
}
\end{table}

\section{Methodology}
\label{sec:methodology}

This section details \methodname's design, resolving the trade-off between semantic richness and computational feasibility. We present its architecture (§\ref{sec:gleaner-framework}) and three core components: the \textbf{Trace Encoder} (§\ref{sec:trace-encoder}) transforms raw traces into lightweight yet expressive representations; the \textbf{Quota Allocator} (§\ref{sec:quota-allocator}) implements an adaptive, alarm-driven budget strategy; and the \textbf{DPP Selector} (§\ref{sec:dpp-selector}) selects a final subset rich in anomalous signals and semantic diversity.

\subsection{The \methodname Framework}
\label{sec:gleaner-framework}

\begin{figure*}[t]
    \centering
    \includegraphics[width=\textwidth]{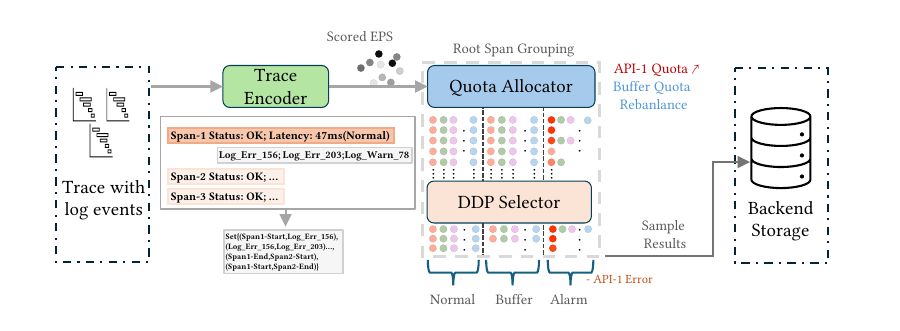}
    \caption{The three-stage pipeline architecture of \methodname. First, the \textbf{Trace Encoder} consumes raw traces to produce semantic representations and anomaly scores. Next, the \textbf{Quota Allocator} adaptively distributes the sampling budget based on real-time monitoring data. Finally, the \textbf{DPP Selector} selects a small, diverse, and anomaly-rich subset of traces.}
    \label{fig:gleaner-architecture}
\end{figure*}

\methodname's overall architecture is depicted in Figure~\ref{fig:gleaner-architecture}.

\begin{table}[t]
\centering
\caption{Key notation and definitions used in this paper.}
\label{tab:notation}
\resizebox{0.7\columnwidth}{!}{%
\begin{tabular}{ll}
\toprule
\textbf{Symbol} & \textbf{Definition} \\
\midrule
$\mathcal{T}$ & A distributed trace, represented as a set of spans \\
$s_i \in \mathcal{T}$ & The $i$-th span in trace $\mathcal{T}$ \\
$L(s_i)$ & The set of log events associated with span $s_i$ \\
$L_w(s_i), L_e(s_i)$ & Subsets of $L(s_i)$ with WARN and ERROR levels, respectively \\
$D(s_i)$ & The duration of span $s_i$ \\
$D_{p90}(G_j)$ & The 90th percentile duration for historical traces in group $G_j$ \\
$E(\mathcal{T})$ & The EPS (event-pair set) representation of trace $\mathcal{T}$ \\
$A(\mathcal{T})$ & The anomaly score of trace $\mathcal{T}$ \\
$\mathcal{B}$ & The total sampling budget (number of traces to select) \\
$\mathcal{G}$ & The set of root span groups: $\{G_1, G_2, \ldots, G_k\}$ \\
$G_j$ & A group of traces sharing the same root span (API endpoint) \\
$q_j$ & The sampling quota allocated to group $G_j$ \\
$\mathcal{A}$ & The set of alerted APIs from external monitoring systems \\
$\mathbf{L}$ & The DPP kernel matrix used by the DPP Selector \\
$\mathcal{S}$ & The final sampled subset of traces output by \methodname \\
\bottomrule
\end{tabular}
}
\end{table}

\subsection{Trace Encoder: A Lightweight Semantic Representation}
\label{sec:trace-encoder}

Inspired by the event-centric view adopted by multimodal RCA systems such as Nezha~\cite{yu2023nezha} and iTCRL~\cite{tian-itcrl-2024}, we re-conceptualize a trace from a graph of spans into a \textbf{set of event pairs (EPS)} that captures both inter-span calls and intra-span log details. Span lifecycle markers and salient in-span events are treated as the basic diagnostic units, but instead of organizing them in a graph, \methodname encodes them as local event pairs for online similarity computation. This representation is deterministic and robust to asynchronous timing variations, semantically rich enough to encode both structure and log content, yet remains computationally tractable through a lightweight, hashable data structure that enables efficient similarity computation. We define a \textbf{trace pattern} as a unique EPS: two traces sharing the same EPS belong to the same pattern, representing identical system behavior despite potentially different trace IDs or timestamps.

Our encoder converts each trace $\mathcal{T}$ into a tuple $(E(\mathcal{T}), A(\mathcal{T}))$ as outlined in Algorithm~\ref{alg:trace_encoder}. For each span $s_i$, we construct a canonical event sequence and apply bi-gram encoding to capture execution flow. For example, the root span in Figure~\ref{fig:semantic_blind_spot} with events $\langle$\texttt{start\_basic}, \texttt{log\_e\_156}, \texttt{log\_e\_203}, \texttt{log\_w\_78}, \texttt{end\_basic}$\rangle$ produces intra-span pairs \{(\texttt{start\_basic}, \texttt{log\_e\_156}), (\texttt{log\_e\_156}, \texttt{log\_e\_203}), (\texttt{log\_e\_203}, \texttt{log\_w\_78}), (\texttt{log\_w\_78}, \texttt{end\_basic})\}. Inter-span relationships link parent-end to child-start events, such as (\texttt{end\_basic}, \texttt{start\_route}). The resulting set $E(\mathcal{T})$ automatically deduplicates redundant pairs from parallel calls.
\begin{algorithm}[t]
\small 
\caption{Trace Encoder}
\label{alg:trace_encoder}
\begin{algorithmic}[1]
\Require Raw trace $\mathcal{T}$ with spans and logs
\Ensure EPS $E(\mathcal{T})$, anomaly score $A(\mathcal{T})$
\State $E \gets \emptyset$, $A \gets 0$
\State Compute $A$ via Eq.~\ref{eq:anomaly_score} \Comment{Errors, logs, latency}
\For{$s_i \in \mathcal{T}$} \Comment{Intra-span encoding}
    \State $\mathcal{E}_i \gets [\text{span\_start}, \text{logs}_{\text{sorted}}, \text{status\_error}, \text{perf\_deg}, \text{span\_end}]$
    \State $E \gets E \cup \{(e_k, e_{k+1}) \mid e_k, e_{k+1} \in \mathcal{E}_i\}$ \Comment{Bi-grams}
\EndFor
\For{$(s_p, s_c)$ in parent-child pairs} \Comment{Inter-span linking}
    \State $E \gets E \cup \{(\text{end}_{s_p}, \text{start}_{s_c})\}$
\EndFor
\State \Return $(E, A)$ \Comment{Set $E$ auto-deduplicates}
\end{algorithmic}
\end{algorithm}

\textbf{Event ID Management and Anomaly Scoring.} We maintain an Event Manager that assigns unique integer IDs to different event types. For log events, we apply Drain~\cite{he2017drain} during collection to strip dynamic variables and map recurring messages to stable templates; we then use the template ID as the event ID. This normalization is essential for preventing state-space explosion from high-cardinality log fields while preserving recurring event semantics. Span start, span end, status error, and performance degradation events are dynamically assigned IDs based on the \texttt{service\_name} + \texttt{span\_name} combination they belong to. Concurrently, we calculate a fine-grained anomaly score $A(\mathcal{T})$ for each trace. This score is a weighted sum of indicators from span status, log levels, and performance degradation, as defined in Equation~\ref{eq:anomaly_score}:
\begin{equation}
\label{eq:anomaly_score}
A(\mathcal{T}) = \sum_{s_i \in \mathcal{T}} (w_{err} \cdot \mathbb{I}_{err}(s_i) + w_{lw} \cdot |L_w(s_i)| + w_{le} \cdot |L_e(s_i)|) + A_{perf}(\mathcal{T})
\end{equation}
where $w_{err}, w_{lw}, w_{le}$ are configurable weights (in this paper, we set $w_{err}{=}5$, $w_{lw}{=}1$, $w_{le}{=}2$); $A_{perf}(\mathcal{T})$ is added only if end-to-end latency exceeds $1.2 \times D_{p90}(G_j)$ for the root-span group $G_j$.

\textit{Hyperparameter Robustness.} The exact form of $A(\mathcal{T})$ is not central to our method; its role is to provide a lightweight relevance prior that separates clearly anomalous traces from the healthy majority. In our benchmark evaluation, varying $w_{err}, w_{lw}, w_{le}$ and the latency score term $A_{perf}(\mathcal{T})$ leaves coverage and entropy metrics completely unchanged, with less than 1\% variance only in the capture rate of diagnostically relevant traces. Empirically, our default setting prioritizes hard failures like status errors over soft signals like logs, mirroring standard SRE triage principles. We emphasize that these parameters remain configurable, allowing operators to dynamically shift focus, such as making $A_{perf}(\mathcal{T})$ more aggressive to ``zoom in'' on performance regressions during tuning campaigns.

\textbf{Intra-Span Canonical Ordering.} For each span $s_i$, we construct a canonical event sequence immune to timing variations common in asynchronous systems. \methodname enforces a deterministic sequence: (1) span start, (2) log events ordered by timestamp, (3) anomaly indicators if present (errors, performance degradation), and (4) span end. We then apply bi-gram encoding to capture the span's internal execution flow as event pairs.

\textbf{Inter-Span Structural Linking.} We extract parent-child span relationships and create event pairs linking the parent's \texttt{span\_end} event to the child's \texttt{span\_start} event. Rather than building a complete call tree, we use this pair-based approach for robustness against \textit{broken traces} common in production, where some spans may have missing parent references due to instrumentation gaps or collection failures.

\textbf{Set-Based Representation.} All generated event pairs are collected into a single set, $E(\mathcal{T})$. This set-based representation is a critical design choice, as it automatically deduplicates redundant pairs that arise from fan-out parallel calls, a common scenario in microservices.

\begin{mybox}
\textbf{Insight 1:} Modeling a trace as a hashable EPS captures both inter-span structure and intra-span semantics while eliminating parallel duplication and enabling efficient Jaccard similarity calculation.
\end{mybox}

\subsection{Quota Allocator: Alarm-Driven Adaptive Budgeting}
\label{sec:quota-allocator}

Production samplers must respond to incidents like SREs do, but existing methods are disconnected from external monitoring, fail to adapt during traffic drops, and risk over-sampling during alarm storms~\cite{xieshuaiyu-tracepicker-2025}. Our allocator integrates with monitoring systems to dynamically adjust sampling budgets, scaling up when failures cause traffic drops while using caps to prevent any single alarm from exhausting resources.

\begin{figure}[t]
    \centering
    \includegraphics[width=0.6\columnwidth]{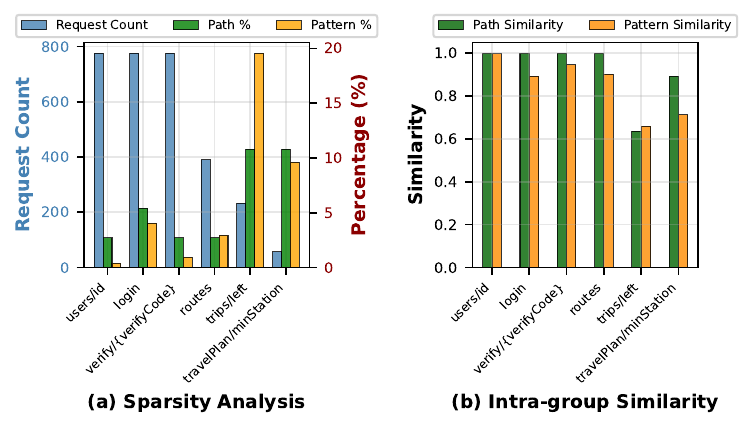}
    \caption{Root span grouping validation on Train-Ticket benchmark. (a) Sparsity analysis: ratio between request volume (QPM) and contribution to total unique paths/patterns for selected API endpoints. (b) Intra-group similarity: average Jaccard similarity of paths and event-pair patterns within each group. High intra-group similarity validates homogeneity for efficient DPP optimization.}
    \label{fig:grouping_validation}
\end{figure}

\textbf{Alarm Generation and Grouping Strategy.}
\methodname groups traces at the \textbf{root span} level, where the root span is the first span of a trace representing the entry point API. This grouping aligns with how production monitoring organizes SLIs for latency and error rate. Root spans provide a stable semantic anchor while enabling efficient downstream processing: endpoints exhibit heterogeneous diversity yet maintain high intra-group homogeneity, as validated in Figure~\ref{fig:grouping_validation}.

We deliberately avoid heavyweight GNN-based clustering, which incurs prohibitive overhead for online streams, and structural hashing like path codes~\cite{xieshuaiyu-tracepicker-2025}, which fragments quota allocations due to spurious groups from minor control-flow variations. Figure~\ref{fig:grouping_validation} shows that while endpoints contribute heterogeneously to total unique paths, they exhibit very high intra-group similarity with path similarity often exceeding 0.9. This confirms root-span grouping correctly identifies the primary semantic axis while filtering artificial structural diversity, producing compact, homogeneous groups required for efficient DPP selection.

\textbf{Two-Layer Allocation Process.}
\methodname maintains a time-based \textit{buffer} that continuously stores recent traces. When an external alarm is triggered, this buffer provides the \textit{normal period} baseline, while incoming traces during the active alarm constitute the \textit{abnormal period}. Our allocation mechanism operates as a two-layer process over these two periods.

\textit{Layer 1: Global Budget Adjustment.} First, we manage the total sampling budget $\mathcal{B}$ to adapt to real-time traffic conditions. Severe system failures often cause sharp traffic volume (QPM) drops due to cascading effects, requiring budget scaling to preserve coverage. When an alarm is triggered, we compare the QPM of the current abnormal period to the historical average from the buffer. If a significant drop is detected, a scaling factor is applied to temporarily increase the total budget $\mathcal{B}$. This adjusted budget is then allocated between the normal and abnormal periods based on their relative QPM, prioritizing the incident period.

\textit{Layer 2: Intra-Group Allocation.} Second, the budget for each period is allocated to the root span groups $G_j \in \mathcal{G}$ within it. For the normal period, the budget is distributed evenly across all groups to maximize baseline diversity and maintain comprehensive system visibility. For the abnormal period, we mimic an SRE's focus: groups $G_j$ corresponding to alerted APIs in $\mathcal{A}$ receive a \textbf{boosted quota} $q_j$, potentially up to 3x the average. This boost is subject to a \textbf{cap} (e.g., 50\% of the period's budget) to prevent any single issue from exhausting resources. The remaining budget is then distributed among non-alarmed groups, guaranteeing baseline coverage for all services (detailed coverage metrics in §\ref{sec:evaluation}). When no alarms are active, the allocator processes the buffer uniformly, preserving system-wide visibility without alarm-driven prioritization.

\begin{mybox}
\textbf{Insight 2:} An effective online sampler should \textbf{mimic an SRE's focus during an incident}. By integrating with external alarms and dynamically rebalancing budget to compensate for traffic drops, the sampler can prioritize the most relevant data when it is needed most.
\end{mybox}

\subsection{DPP Selector: Optimized Selection for Diversity and Relevance}
\label{sec:dpp-selector}

We employ Determinantal Point Processes (DPP)~\cite{chen_fast_2018} to select diverse, high-quality trace subsets. The DPP kernel matrix $\mathbf{L}$ encodes anomaly scores $A(\mathcal{T})$ on its diagonal and EPS Jaccard similarities on off-diagonal entries. Standard DPP is computationally expensive; we introduce two optimizations for online deployment.

\textbf{Early Termination via High Intra-Group Similarity.} Root-span grouping creates semantically coherent groups by construction—all traces in group $G_j$ originate from the same API endpoint. This homogeneity (Figure~\ref{fig:grouping_validation}b shows intra-group Jaccard similarity often exceeding 0.9) enables fast greedy DPP~\cite{chen_fast_2018} to converge rapidly, as marginal diversity gains drop below threshold $\epsilon$ after selecting few candidates.

\textbf{Persistent Cross-Batch Similarity Caching.} EPS patterns exhibit high temporal stability—the same patterns recur as users repeatedly invoke API endpoints. We implement a global LRU cache keyed by $\text{hash}(E(\mathcal{T}_i), E(\mathcal{T}_j))$ that persists across batches. This achieves a 91.8\% hit rate, dramatically reducing redundant Jaccard similarity calculations with negligible memory overhead.

\begin{mybox}
\textbf{Insight 3:} The prohibitive cost of DPPs can be overcome by exploiting the \textbf{high intra-group similarity unique to our design}. Unlike post-hoc clustering methods, our root-span grouping produces semantically coherent groups by construction, enabling fast greedy DPP~\cite{chen_fast_2018} to converge rapidly. Combined with persistent cross-batch similarity caching, this makes DPP practical for high-throughput online streams.
\end{mybox}

These optimizations make DPP practical for high-throughput online streams, enabling diversity-aware selection at sub-millisecond latency.

\section{Evaluation}
\label{sec:evaluation}

We evaluate \methodname to answer four key research questions:

\begin{itemize}
    \item \textbf{RQ1:} How does \methodname compare to state-of-the-art samplers in sampling quality, diversity, and anomaly capture?
    \item \textbf{RQ2:} What is the contribution of each design component in \methodname's overall performance?
    \item \textbf{RQ3:} Does \methodname improve downstream root cause analysis (RCA) accuracy?
    \item \textbf{RQ4:} What is \methodname's efficiency in terms of runtime overhead, budget control, and information density?
\end{itemize}
\subsection{Experimental Setup}

\textbf{Benchmark Datasets.}
To ensure a comprehensive and robust evaluation, we use two distinct sets of benchmarks, which we denote as Dataset $\mathcal{A}$ and Dataset $\mathcal{B}$.

\textit{1) In-depth Fault Analysis Dataset (Dataset $\mathcal{A}$):} Our primary evaluation is performed on a new, large-scale dataset we generated from the Train-Ticket~\cite{trainticket-paper} benchmark. To rigorously test sampler performance under failure scenarios, we collected 161 fault injection cases using ChaosMesh~\cite{ChaosMesh2025}, covering 16 types of production-realistic failures across application, network, HTTP, and container layers. This dataset, \textbf{Dataset $\mathcal{A}$}, allows for in-depth analysis of anomaly capture and its impact on downstream RCA.

\textit{2) Cross-System Generalization Dataset (Dataset $\mathcal{B}$):} To evaluate \methodname's generalizability, we use a public dataset from TracePicker~\cite{xieshuaiyu-tracepicker-2025}. This collection, \textbf{Dataset $\mathcal{B}$}, contains traces from five different well-known microservice systems, allowing us to assess performance across diverse architectures and workloads. Since these datasets consist only of traces without corresponding logs or alarm labels, we use a variant of \methodname (\methodname w/o Logs w/o Alarms) for this part of the evaluation.

\begin{figure*}[t]
    \centering
    \begin{minipage}{0.37\textwidth}
        \centering
        \footnotesize
        \captionsetup{type=table}
        \caption{Distribution of fault injection cases in Dataset $\mathcal{A}$.}
        \label{tab:fault_distribution}
        \resizebox{\linewidth}{!}{%
        \begin{tabular}{l|cccc}
        \toprule
        & \textbf{App} & \textbf{Network} & \textbf{HTTP} & \textbf{Container} \\
        \midrule
        \textbf{Type} & CPU/mem & Delays, loss, & Req delays, & Container \\
        & code & partitions & aborts, tamper & kills \\
        \midrule
        \textbf{Count} & 46 & 39 & 58 & 18 \\
        \bottomrule
        \end{tabular}%
        }
    \end{minipage}%
    \hfill
    \begin{minipage}{0.60\textwidth}
        \centering
        \footnotesize
        \captionsetup{type=table}
        \caption{Statistics of the benchmark datasets used in our evaluation.}
        \label{tab:dataset_stats}
        \resizebox{\linewidth}{!}{%
        \begin{tabular}{l|c|ccccc}
        \toprule
        \multirow{2}{*}{\textbf{Metric}} & \textbf{Dataset $\mathcal{A}$} & \multicolumn{5}{c}{\textbf{Dataset $\mathcal{B}$ (from TracePicker)}} \\
        \cmidrule{2-7}
        & \textbf{(Ours)} & \textbf{Train-Ticket} & \textbf{Media} & \textbf{OnlineBoutique} & \textbf{Sock-Shop} & \textbf{Social-Network} \\
        \midrule
        Total Traces & 1,474,537 & 22,000 & 28,710 & 42,074 & 43,472 & 32,123 \\
        Total Trace Patterns & 285,059 & 2,937 & 3,841 & 4,216 & 271 & 2,443 \\
        Total Entries & 24 & 16 & 2 & 6 & 8 & 1 \\
        \bottomrule
        \end{tabular}%
        }
    \end{minipage}
\end{figure*}

\textbf{Baseline Methods.}
We compare \methodname against a suite of representative and state-of-the-art samplers:
\begin{itemize}
    \item \textbf{Random}: Uniformly samples traces at a fixed probability.
    \item \textbf{Sieve}~\cite{huang-sieve-2021}: An online sampler that prioritizes structurally and temporally rare traces using Robust Random Cut Forest (RRCF).
    \item \textbf{Sifter}~\cite{las-casas-sifter-2019}: An online sampler that models common trace structures and prioritizes those that deviate from the model.
    \item \textbf{TraStrainer}~\cite{huang2024trastrainer}: An adaptive sampler that correlates traces with external system metrics. We provide it with key metrics (CPU, memory, P50/P90 latency) as input. We also include a \textbf{TraStrainer w/o Metrics} variant to assess its trace-only performance.
    \item \textbf{TracePicker}~\cite{xieshuaiyu-tracepicker-2025}: A state-of-the-art online sampler that uses an optimization-based approach to prioritize anomalous traces and then maximize path coverage.
\end{itemize}

\textbf{Implementation Details.}
All experiments were run on AMD EPYC 9754 128-Core Processors (2 cores per sampler). For samplers with imprecise budget control, we apply standard normalization~\cite{las-casas-sifter-2019,huang2024trastrainer}. For baseline methods, we use publicly released code where available, and for methods without public implementations, we contacted the related authors and strictly followed their published specifications. All metrics are computed independently for each of the 161 fault injection cases and then averaged to reduce variance from stochastic factors.

\subsection{Evaluation Metrics}

We evaluate \methodname's performance across two dimensions: intrinsic sampling quality in RQ1, RQ2, and RQ4, and downstream task effectiveness in RQ3.

For sampling quality, we adopt established metrics from prior work~\cite{he2023steam,tian-itcrl-2024,huang2024trastrainer,xieshuaiyu-tracepicker-2025}. Following prior work, we organize them into three complementary families: \emph{coverage} measures how much of the original observability space remains after sampling, \emph{entropy} measures whether the sampled traces preserve sufficient information density rather than collapsing onto a few dominant patterns, and \emph{proportion} measures whether the sampler is biased toward diagnostically valuable traces.
\begin{itemize}[leftmargin=*]
    \item \textbf{Coverage metrics:} We report \textbf{API Coverage}, \textbf{Path Coverage}, and \textbf{Trace Pattern Coverage}. API Coverage measures whether the sampled set preserves visibility across business entry points. Path Coverage measures whether distinct execution flows are retained; in our implementation, this uses a deduplicated path encoding, where exact parallel duplicates at the same level are removed before path comparison. Trace Pattern Coverage (as defined in §\ref{sec:trace-encoder}) measures how many EPS-level behavioral patterns remain after sampling.
    \item \textbf{Entropy metric:} We report \textbf{Shannon Entropy} to quantify whether sampled traces are broadly distributed across patterns, rather than being concentrated in a few dominant behaviors.
    \item \textbf{Proportion metrics:} We report \textbf{Proportion Anomaly} and \textbf{Proportion Rare} to quantify whether the sampler preferentially preserves diagnostically valuable traces from the anomaly and long-tail perspectives, respectively.
    \item \textbf{Downstream utility metric:} For RCA evaluation, we use \textbf{Accuracy@k}, the proportion of cases where the true root cause is ranked in the top-k results.
\end{itemize}

For efficiency analysis, we measure \textbf{Runtime Per Trace} and \textbf{Actual Sampling Rate}. We also introduce the \textbf{Benefit-Cost Ratio} ($BCR$) to quantify the efficiency of unique pattern discovery:
\begin{equation}
    BCR = \frac{N_{unique}}{N_{sample}}
\end{equation}
where $N_{unique}$ is the number of unique trace patterns discovered and $N_{sample}$ is the actual count of sampled traces. A higher $BCR$ indicates a more efficient discovery of diverse trace patterns within the sampling budget.

\subsection{RQ1: Sampling Quality and Diversity}
\label{sec:rq1}

This experiment assesses \methodname's intrinsic sampling quality across three evaluation scenarios: coverage and diversity on Dataset $\mathcal{A}$, anomaly and rarity capture on Dataset $\mathcal{A}$, and cross-system generalization on Dataset $\mathcal{B}$.

\begin{figure}[t]
    \centering
    \includegraphics[width=0.8\columnwidth]{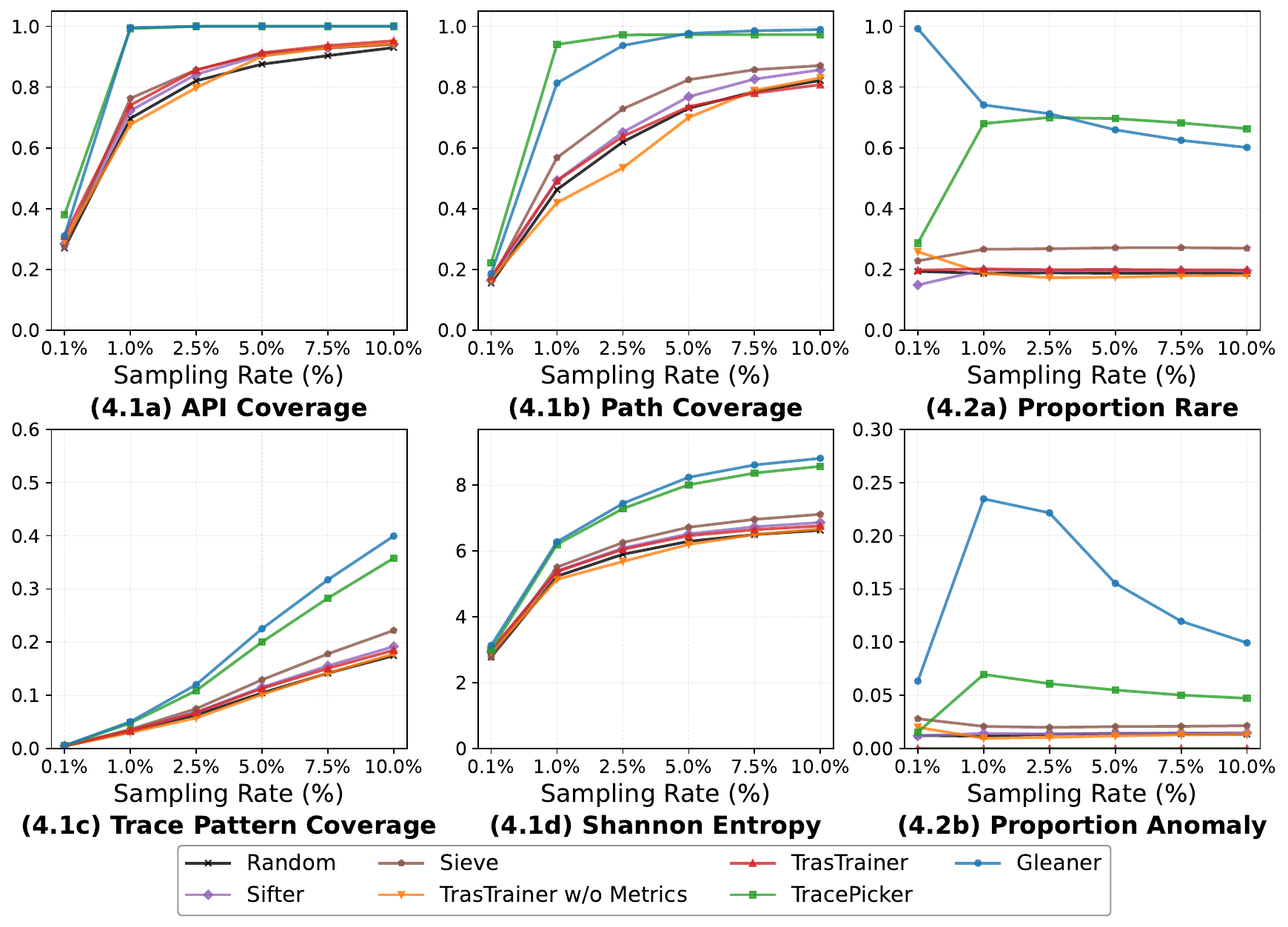}
    \caption{Sampling quality evaluation on Dataset $\mathcal{A}$. Left two columns (4.1a-d): Coverage and diversity metrics showing \methodname's consistent superiority across all dimensions. Rightmost column (4.2a-b): Anomaly and rarity capture demonstrating \methodname's dramatic advantage in prioritizing diagnostically relevant traces.}
    \label{fig:rq1_quality}
\end{figure}

\textbf{Coverage and Diversity on Dataset $\mathcal{A}$.}
Figure~\ref{fig:rq1_quality}1a-d shows \methodname's superior performance across coverage and diversity metrics. At 10\% sampling rate, \methodname achieves 11.6\%\textasciitilde{}128.7\% improvement in Trace Pattern Coverage and 2.8\%\textasciitilde{}32.9\% gains in Shannon Entropy over baselines. At very low rates ($\leq 1\%$), TracePicker achieves slightly better Path and API Coverage due to its optimization focus under severe constraints, but \methodname dominates at practical rates ($\geq 1\%$).

\textbf{Anomaly and Rarity Capture on Dataset $\mathcal{A}$.}
Figure~\ref{fig:rq1_quality}.2a-b shows \methodname's advantage in capturing anomalous and rare traces. At 0.1\% sampling rate, \methodname achieves Proportion Rare of 0.992, 3.5$\times$ higher than TracePicker. For Proportion Anomaly, \methodname maintains 2.1\textasciitilde{}8.3$\times$ improvement over baselines across all rates, validating alarm-driven quota allocation.

\begin{figure}[t]
    \centering
    \includegraphics[width=0.8\columnwidth]{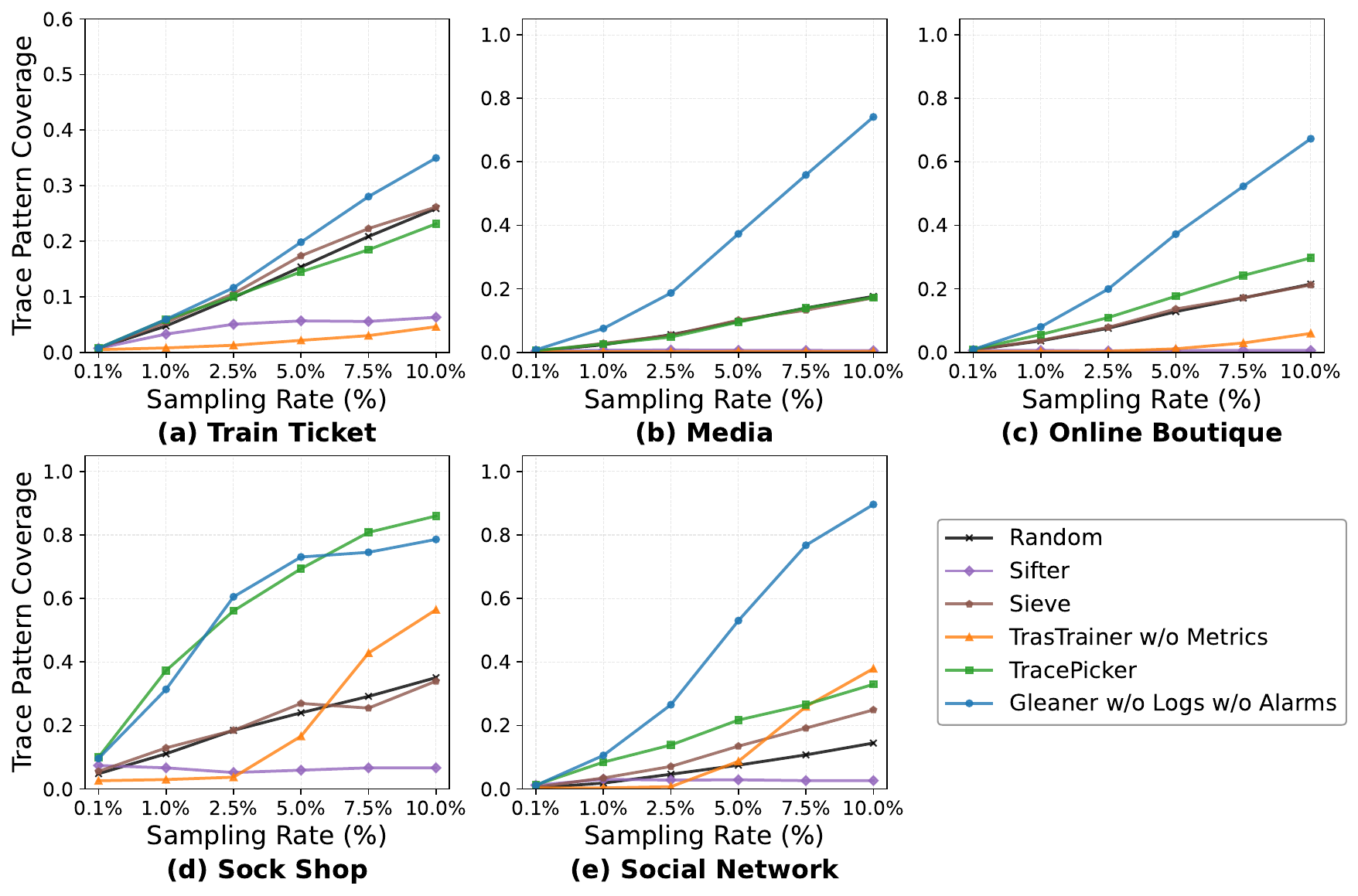}
    \caption{Cross-system evaluation on Dataset $\mathcal{B}$ (5 microservice benchmarks). \methodname consistently outperforms baselines across most systems, with comparable performance to TracePicker on simpler architectures.}
    \label{fig:cross_system_coverage}
\end{figure}

\textbf{Cross-System Generalization on Dataset $\mathcal{B}$.}
Figure~\ref{fig:cross_system_coverage} shows \methodname (w/o Logs w/o Alarms) consistently outperforms baselines across five microservice benchmarks. At 10\% sampling rate, \methodname achieves average Trace Pattern Coverage of 0.689, outperforming TracePicker by 0.088 and other methods by 2.8\textasciitilde{}4.0$\times$. The main exception is SockShop, whose workload is dominated by a single front-end entry point. Because \methodname's allocator groups traces by root span, this architecture largely removes the grouping advantage and makes the method behave closer to a structure-driven diversity sampler; even under this unfavorable condition, \methodname remains competitive with TracePicker.

\begin{mybox}
\textbf{Finding 1:} \methodname produces a sample set that is more comprehensive, diverse, and diagnostically relevant than state-of-the-art baselines on Dataset $\mathcal{A}$, and generalizes well across systems in Dataset $\mathcal{B}$. Its gains are largest in systems with multiple business entry points and richer in-span events; when those conditions weaken, as in SockShop, \methodname degrades gracefully toward a structure-driven diversity sampler while remaining competitive.
\end{mybox}

\subsection{RQ2: Ablation Study}
\label{sec:rq2}

To understand the contribution of \methodname's core components and design choices, we conduct an ablation study. Table~\ref{tab:ablation_variants} summarizes all variants. We report results in two groups: Group~1 examines the impact of data sources and representation methods, and Group~2 explores the balance between sampling strategies. All variants are evaluated on Dataset $\mathcal{A}$ using the same metrics as RQ1.

\begin{table*}[t]
  \caption{Summary of \methodname variants designed for ablation study. The table disentangles the contribution of input signals, representation methods, and sampling strategies.}
  \label{tab:ablation_variants}
  \resizebox{\textwidth}{!}{%
  \begin{tabular}{l|cc|c|ccc|l}
    \toprule
    \multirow{2}{*}{\textbf{Variant Name}} & \multicolumn{2}{c|}{\textbf{Input Signals}} & \textbf{Representation} & \multicolumn{3}{c|}{\textbf{Sampling Strategy}} & \multirow{2}{*}{\textbf{Motivation / Focus}} \\
    \cmidrule(lr){2-3} \cmidrule(lr){4-4} \cmidrule(lr){5-7}
     & \textbf{Logs} & \textbf{Alarms} & \textbf{Structure Enc.} & \textbf{Grouping} & \textbf{Anomaly} & \textbf{Diversity} & \\
    \midrule
    \textbf{\methodname} & \checkmark & \checkmark & \textbf{EPS} & \checkmark & \checkmark & \checkmark & \textbf{Proposed Method (Full)} \\
    \midrule
    \textit{Group 1: Inputs \& Rep.} & & & & & & & \\
    \textbf{w/o Logs} &  & \checkmark & EPS & \checkmark & \checkmark & \checkmark & Impact of semantic richness \\
    \textbf{w/o Alarms} & \checkmark &  & EPS & \checkmark & \checkmark & \checkmark & Impact of dynamic quota \\
    \textbf{w/o Logs \& Alarms} &  &  & EPS & \checkmark & \checkmark & \checkmark & Baseline structural performance \\
    \textbf{WL Kernel} & \checkmark & \checkmark & \textbf{Graph} & \checkmark & \checkmark & \checkmark & Efficiency check (EPS vs. Graph) \\
    \midrule
    \textit{Group 2: Strategies} & & & & & & & \\
    \textbf{Pure Diversity} & \checkmark & \checkmark & EPS &  &  & \checkmark & Impact of Grouping \& Anomaly \\
    \textbf{Pure Anomaly} & \checkmark & \checkmark & EPS &  & \checkmark &  & Impact of Grouping \& Diversity \\
    \textbf{w/o Anomaly} & \checkmark & \checkmark & EPS & \checkmark &  & \checkmark & Impact of Anomaly prioritization \\
    \textbf{w/o Diversity} & \checkmark & \checkmark & EPS & \checkmark & \checkmark &  & Impact of Diversity selection \\
    \bottomrule
  \end{tabular}%
  }
\end{table*}

\begin{figure}[t]
    \centering
    \includegraphics[width=\columnwidth]{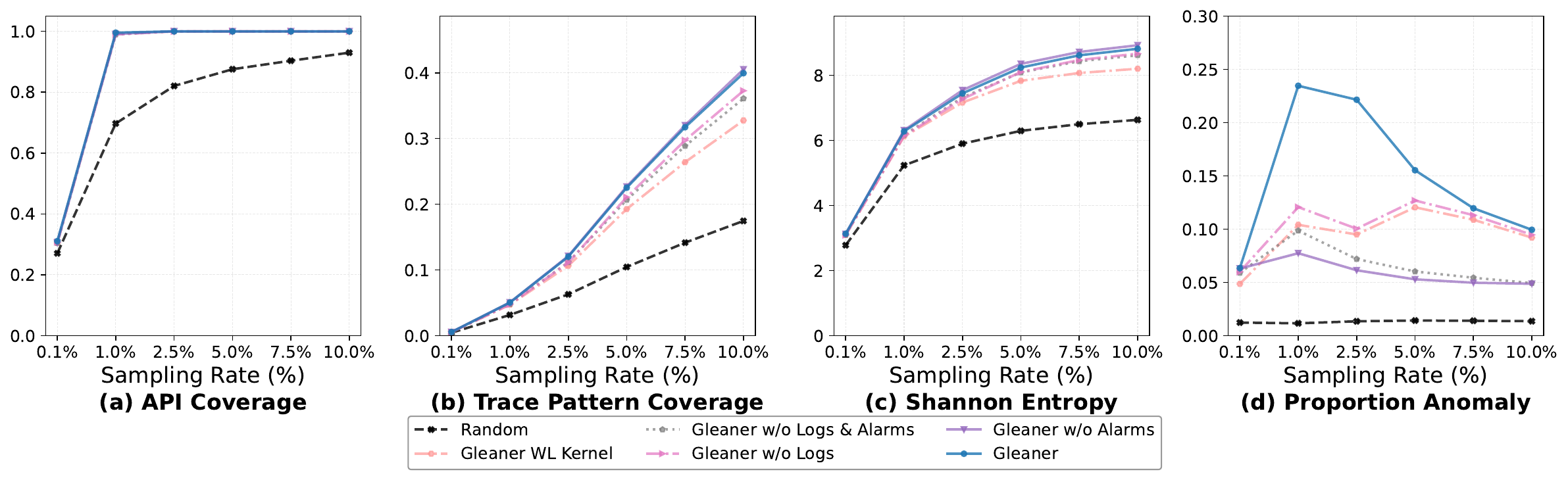}
    \caption{Ablation study Group 1: Impact of semantic components (logs, alarms) and structural representation (vs. WL Kernel) on Dataset $\mathcal{A}$.}
    \label{fig:ablation_group1}
\end{figure}

\textbf{Group 1: Component Effectiveness (Figure~\ref{fig:ablation_group1}).}
This group validates the impact of input signals and structural representation.

Overall, Group~1 shows that the structural signal captured by our Event Pair Set (EPS) representation provides a strong foundation, while logs and alarms improve fault relevance. Even without logs and alarms, the sampler remains consistently better than Random across the quality metrics, indicating that EPS alone already captures useful execution regularities. Counter-intuitively, replacing EPS with the WL Graph Kernel yields \textit{lower} diversity: pattern coverage drops from 39.9\% to 32.8\% while runtime increases 8.4$\times$ (quantified in Table~\ref{tab:efficiency}). This suggests that exact structural matching is too rigid for microservices, where concurrent fan-out patterns create structurally distinct graphs for identical logical requests; EPS is more robust to these concurrency-induced variations. Finally, disabling alarms reveals an expected trade-off: proportion anomaly drops significantly, while entropy and coverage metrics slightly increase. This occurs because without alarm-driven quota allocation, the budget distributes more uniformly across all API groups than concentrating on alarm groups, leading to marginally higher diversity at the cost of reduced fault relevance.

\begin{figure}[t]
    \centering
    \includegraphics[width=\columnwidth]{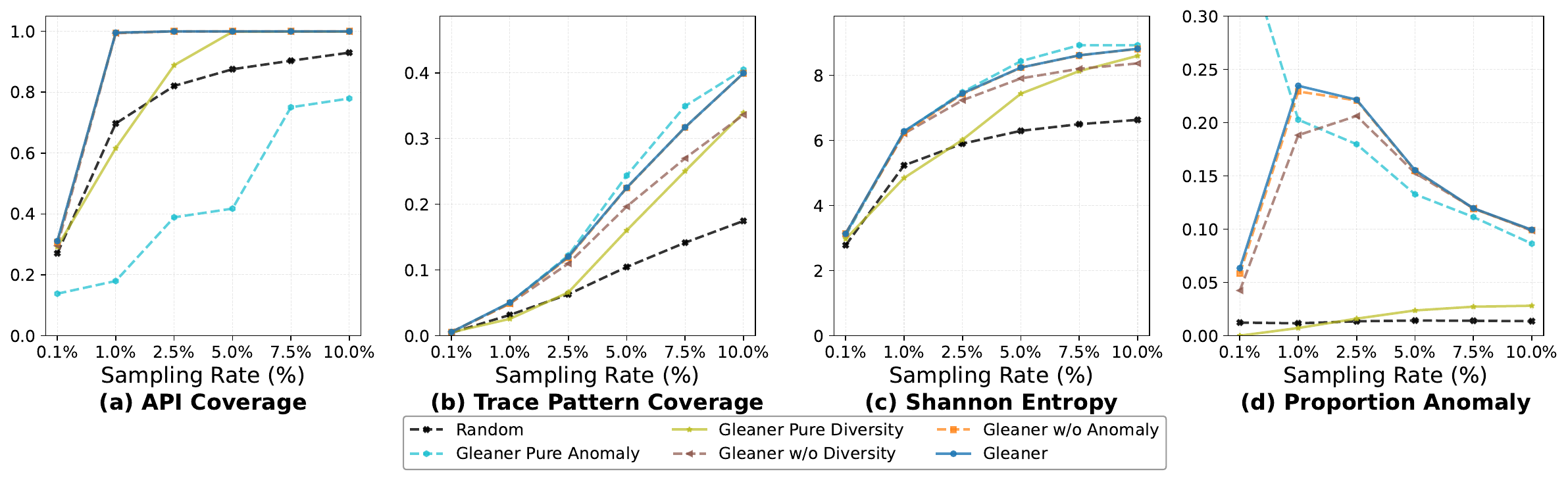}
    \caption{Ablation study Group 2: Performance comparison of different sampling strategies, highlighting the synergy between diversity and anomaly scoring on Dataset $\mathcal{A}$.}
    \label{fig:ablation_group2}
\end{figure}

\textbf{Group 2: Strategy Balance (Figure~\ref{fig:ablation_group2}).}
This group explores the necessity of balancing anomaly prioritization with diversity by examining extreme strategies.

Group~2 demonstrates that each component is necessary: removing any single mechanism degrades the corresponding quality dimension. Most notably, \textit{Pure Anomaly} exhibits an interesting trade-off: while it achieves higher entropy and pattern coverage than the full model, API coverage collapses at low rates (1\%: 0.18 vs Random's 0.70); without grouping, the sampler fixates on whichever endpoints currently exhibit anomalies in each batch, leading to highly unbalanced API representation. Meanwhile, \textit{Pure Diversity} shows that removing root-span grouping causes coverage to drop significantly: without per-group batching, different batches repeatedly select the same globally diverse traces, reducing overall coverage. The intermediate variant \textit{w/o Diversity} maintains most metrics but suffers reduced pattern representativeness, while \textit{w/o Anomaly} performs nearly identically to the full model except for slightly lower anomaly capture. Overall, the full \textit{\methodname} achieves the optimal balance across all dimensions.

\begin{mybox}
\textbf{Finding 2:} Ablations confirm the role of each design component. EPS provides a strong structural foundation and is more effective than the Graph Kernel alternative. Logs and alarms improve fault relevance, and removing alarms induces the expected trade-off between anomaly focus and diversity. Root-span grouping, anomaly prioritization, and diversity selection are all necessary to achieve optimal sampling quality.
\end{mybox}

\subsection{RQ3: Impact on Downstream Root Cause Analysis}
\label{sec:rq3}

We evaluate how sampler choice affects automated Root Cause Analysis (RCA) accuracy using three open-source RCA algorithms: \textbf{MicroRCA}~\cite{wu2020microrca}, \textbf{Nezha}~\cite{yu2023nezha}, and \textbf{ShapleyIQ}~\cite{li2023shapleyiq}. For fair comparison, we provide aggregated metrics (p90 latency, error count) to supporting tools. Results are in Table~\ref{tab:rca_results}.

\begin{figure*}[t]
    \centering
    \begin{minipage}{0.56\textwidth}
        \centering
        \footnotesize
        \captionsetup{type=table}
        \caption{RCA accuracy comparison on Dataset $\mathcal{A}$. Unsampled row shows performance using all traces. \methodname achieves the highest accuracy across all tools.}
        \label{tab:rca_results}
        \resizebox{\linewidth}{!}{%
        \begin{tabular}{l|cc|cc|cc|cc|cc|cc}
        \toprule
        \multirow{3}{*}{\textbf{Sampler}} & \multicolumn{4}{c|}{\textbf{MicroRCA}} & \multicolumn{4}{c|}{\textbf{Nezha}} & \multicolumn{4}{c}{\textbf{ShapleyIQ}} \\
        \cmidrule{2-5} \cmidrule{6-9} \cmidrule{10-13}
        & \multicolumn{2}{c|}{$\mathbf{\mathrm{AC}@1}$} & \multicolumn{2}{c|}{$\mathbf{\mathrm{AC}@3}$} & \multicolumn{2}{c|}{$\mathbf{\mathrm{AC}@1}$} & \multicolumn{2}{c|}{$\mathbf{\mathrm{AC}@3}$} & \multicolumn{2}{c|}{$\mathbf{\mathrm{AC}@1}$} & \multicolumn{2}{c}{$\mathbf{\mathrm{AC}@3}$} \\
        & \textbf{1\%} & \textbf{10\%} & \textbf{1\%} & \textbf{10\%} & \textbf{1\%} & \textbf{10\%} & \textbf{1\%} & \textbf{10\%} & \textbf{1\%} & \textbf{10\%} & \textbf{1\%} & \textbf{10\%} \\
        \midrule
        Random & 0.39 & 0.45 & 0.44 & 0.49 & 0.06 & 0.09 & 0.15 & 0.19 & 0.17 & 0.40 & 0.30 & 0.56 \\
        TracePicker & 0.45 & 0.45 & 0.49 & 0.49 & 0.08 & 0.06 & 0.19 & 0.20 & 0.26 & 0.39 & 0.45 & 0.48 \\
        Sieve & 0.42 & 0.45 & 0.47 & 0.49 & 0.11 & 0.09 & 0.23 & 0.17 & 0.27 & 0.47 & 0.40 & 0.60 \\
        Sifter & 0.41 & 0.45 & 0.48 & 0.49 & 0.02 & 0.08 & 0.15 & 0.25 & 0.18 & 0.40 & 0.37 & 0.58 \\
        TraStrainer & 0.40 & 0.45 & 0.49 & 0.49 & 0.10 & 0.07 & 0.24 & 0.19 & 0.06 & 0.03 & 0.19 & 0.22 \\
        TraStrainer w/o M & 0.38 & 0.45 & 0.45 & 0.50 & 0.09 & 0.09 & 0.26 & 0.24 & 0.09 & 0.32 & 0.23 & 0.57 \\
        \midrule
        \textbf{\methodname} & \textbf{0.45} & \textbf{0.45} & \textbf{0.50} & \textbf{0.50} & \textbf{0.19} & \textbf{0.14} & \textbf{0.45} & \textbf{0.35} & \textbf{0.56} & \textbf{0.52} & \textbf{0.64} & \textbf{0.61} \\
        \midrule
        Unsampled & \multicolumn{2}{c|}{0.45} & \multicolumn{2}{c|}{0.50} & \multicolumn{2}{c|}{0.11} & \multicolumn{2}{c|}{0.27} & \multicolumn{2}{c|}{0.41} & \multicolumn{2}{c}{0.57} \\
        \bottomrule
        \end{tabular}%
        }
    \end{minipage}%
    \hfill
    \begin{minipage}{0.425\textwidth}
        \centering
        \footnotesize
        \captionsetup{type=table}
        \caption{Ablation analysis on RCA accuracy. The synergistic combination of anomaly, grouping, and diversity outperforms any single strategy.}
        \label{tab:ablation_rca}
        \resizebox{\linewidth}{!}{%
        \begin{tabular}{l|cc|cc|cc|cc}
        \toprule
        \multirow{3}{*}{\textbf{Sampler}} & \multicolumn{4}{c|}{\textbf{Nezha}} & \multicolumn{4}{c}{\textbf{ShapleyIQ}} \\
        \cmidrule{2-9}
        & \multicolumn{2}{c|}{$\mathbf{\mathrm{AC}@1}$} & \multicolumn{2}{c|}{$\mathbf{\mathrm{AC}@3}$} & \multicolumn{2}{c|}{$\mathbf{\mathrm{AC}@1}$} & \multicolumn{2}{c}{$\mathbf{\mathrm{AC}@3}$} \\
        & \textbf{1\%} & \textbf{10\%} & \textbf{1\%} & \textbf{10\%} & \textbf{1\%} & \textbf{10\%} & \textbf{1\%} & \textbf{10\%} \\
        \midrule
        Pure Anomaly & 0.03 & 0.13 & 0.12 & 0.24 & 0.35 & 0.49 & 0.43 & 0.59 \\
        Pure Diversity & 0.05 & 0.09 & 0.14 & 0.22 & 0.11 & 0.47 & 0.21 & 0.58 \\
        w/o Diversity & 0.10 & 0.07 & 0.23 & 0.19 & 0.45 & 0.49 & 0.60 & 0.60 \\
        w/o Anomaly & 0.14 & 0.11 & 0.29 & 0.24 & 0.57 & 0.52 & 0.65 & 0.60 \\
        w/o Alarms & 0.12 & 0.13 & 0.32 & 0.31 & 0.42 & 0.45 & 0.58 & 0.60 \\
        w/o Logs & 0.13 & 0.11 & 0.32 & 0.27 & 0.50 & 0.51 & 0.62 & 0.62 \\
        w/o Logs w/o Alarms & 0.14 & 0.13 & 0.28 & 0.30 & 0.45 & 0.48 & 0.58 & 0.61 \\
        \midrule
        \textbf{\methodname} & \textbf{0.19} & \textbf{0.14} & \textbf{0.45} & \textbf{0.35} & \textbf{0.56} & \textbf{0.52} & \textbf{0.64} & \textbf{0.61} \\
        \midrule
        Unsampled & \multicolumn{2}{c|}{0.11} & \multicolumn{2}{c|}{0.27} & \multicolumn{2}{c|}{0.41} & \multicolumn{2}{c}{0.57} \\
        \bottomrule
        \end{tabular}%
        }
    \end{minipage}
\end{figure*}

Table~\ref{tab:rca_results} shows \methodname's substantial impact on RCA accuracy. At 1\% sampling rate, \methodname achieves +107\% for Nezha ($\mathrm{AC}@1$: 0.19 vs. 0.11), +42\% for Nezha ($\mathrm{AC}@3$: 0.45 vs. 0.26), and +107\% for ShapleyIQ ($\mathrm{AC}@1$: 0.56 vs. 0.27) compared to next-best samplers.

Most strikingly, \methodname's 1\% sample enables RCA tools to \textit{outperform the complete unsampled dataset}: ShapleyIQ achieves +36.6\% $\mathrm{AC}@1$ (0.56 vs. 0.41), Nezha achieves +72.7\% $\mathrm{AC}@1$ (0.19 vs. 0.11) and +66.7\% $\mathrm{AC}@3$ (0.45 vs. 0.27), while MicroRCA maintains equivalent accuracy with 99\% less data. This reveals that strategic curation enhances signal quality beyond quantity by concentrating diagnostically-rich signals and filtering redundant traces.

\textit{Why does sampling outperform full data?} This result runs counter to the intuition that "more data is better." We investigated this by expanding our evaluation to over 10 additional RCA algorithms. We found that algorithms like MicroRCA were notably insensitive to sampling strategies. This insensitivity arises because MicroRCA locates root causes by analyzing Pearson correlations between aggregated service metrics, such as P90 latency, and system metrics like CPU usage on an attribute graph. Since accurate statistical features are provided as inputs and the service topology is preserved, its Personalized PageRank algorithm remains stable regardless of specific trace selection. In contrast, other tools like DiagFusion~\cite{zhang2023robust} and EADRO~\cite{lee2023eadro} failed to localize faults in our complex dataset regardless of data volume, achieving less than 10\% accuracy. Thus, we present MicroRCA as a stability baseline, while focusing on Nezha and ShapleyIQ to demonstrate the impact of data quality.

To pinpoint the source of this performance gain, we conducted a detailed ablation study on \methodname's components (Table~\ref{tab:ablation_rca}). We first observe that disjoint sampling strategies, such as \textit{Pure Anomaly}, \textit{Pure Diversity}, and \textit{w/o Diversity}, consistently fail to surpass the full dataset baseline. In fact, without the synergy of components, these ablated variants often yield results inferior to even simple \texttt{Random} sampling. This indicates that neither anomaly filtering nor structural diversity alone is sufficient; their synergy is essential. Beyond this general failure, the results reveal that different RCA principles demand different data properties.

For causal-inference-based tools like \textbf{ShapleyIQ}, structural integrity is paramount. When we remove grouping in the \textit{Pure Diversity} variant, ShapleyIQ's $\mathrm{AC}@3$ at 1\% sampling rate collapses from 0.64 to 0.21. In contrast, removing anomaly weighting in the \textit{w/o Anomaly} variant has little impact, maintaining a high $\mathrm{AC}@3$ of 0.65 compared to 0.64 for the full model. This behavior occurs because ShapleyIQ relies on comparing normal and abnormal executions to calculate ``marginal contributions.'' By enforcing uniform sampling across API endpoints through grouping, \methodname ensures a complete ``healthy baseline'' for every endpoint, preventing the missing-baseline failures that cripple the ungrouped \textit{Pure Diversity} variant.

Conversely, statistical-pattern-based tools like \textbf{Nezha} depend heavily on anomaly-guided diversity. Disjoint strategies fail dramatically here: for instance, \textit{Pure Anomaly} (0.03), \textit{Pure Diversity} (0.05), and \textit{w/o Diversity} (0.10) all fall far below the \textit{Unsampled} baseline (0.11) in $\mathrm{AC}@1$ at 1\% rate. Notably, \textit{Pure Anomaly} and \textit{Pure Diversity} even underperform \texttt{Random} sampling (0.06). Similarly, the \textit{w/o Anomaly} variant, which provides diversity without biasing, degrades significantly as $\mathrm{AC}@3$ drops from 0.45 to 0.29. This specific sensitivity indicates that Nezha, which contrasts event pattern frequencies, requires a dataset that is both diverse to cover various failure modes and anomaly-biased to enhance the signal-to-noise ratio. Furthermore, the inclusion of \textit{log encoding} captures failures manifested only in logs, providing the necessary multimodal evidence for complex faults.

\begin{mybox}
\textbf{Finding 3:} \methodname consistently improves downstream RCA accuracy and can even outperform using the full dataset at low sampling rates. The gain comes from curating diagnostically rich traces: grouping preserves healthy baselines for causal inference, and anomaly-guided diversity and log encoding strengthen the signal for pattern-based tools.
\end{mybox}

\subsection{RQ4: Efficiency Analysis}
\label{sec:rq4}

This RQ assesses sampler efficiency under realistic online conditions with a 5\% target sampling rate. To accurately reflect real-world deployment scenarios, we do not artificially constrain the budget of non-deterministic samplers, allowing their effective sampling rates to be governed by their own algorithms. The results in Table~\ref{tab:efficiency} reveal significant differences in practical deployability.

\begin{table}[t]
\centering
\caption{Efficiency comparison on Dataset $\mathcal{A}$ at 5\% target sampling rate. \methodname achieves optimal balance across all three efficiency dimensions.}
\label{tab:efficiency}
\resizebox{0.7\columnwidth}{!}{%
\begin{tabular}{l|ccc}
\toprule
\textbf{Sampler} & \textbf{Runtime (ms)} & \textbf{Benefit-Cost Ratio} & \textbf{Actual Rate (\%)} \\
\midrule
\textbf{\methodname} & \textbf{0.741} & \textbf{0.81} & \textbf{5.0} \\
\textbf{\methodname WL Kernel} & 6.251 & 0.73 & \textbf{5.0} \\
TracePicker & 1.008 & 0.76 & \textbf{5.0} \\
Sieve & 0.629 & 0.33 & 33.2 \\
Sifter & 1.220 & 0.21 & 94.3 \\
TraStrainer & 69.857 & 0.43 & 11.2 \\
TraStrainer w/o M & 1.349 & 0.17 & 72.0 \\
Random & 0.008 & 0.41 & \textbf{5.0} \\
\bottomrule
\end{tabular}%
}
\end{table}

\methodname's efficiency profile (Table~\ref{tab:efficiency}) validates the practical viability of our optimizations: persistent cross-batch similarity caching and early termination enable 0.741ms per-trace latency (26.5\% faster than TracePicker, 94$\times$ faster than TraStrainer) while achieving the highest benefit-cost ratio of 0.81. Moreover, \methodname maintains strict budget control with exactly 5.0\% sampling rate, in contrast to non-deterministic samplers that exhibit severe overruns up to 18.86$\times$ over budget.

We additionally profiled the memory footprint of our current Python prototype. Processing 88,266 spans in one run yielded a peak memory usage of approximately 1.3GB. Given the data volume involved, this result suggests that the prototype-level implementation remains within a practical range for a dedicated sampler process.

The comparison with \textit{\methodname WL Kernel} directly validates our EPS design choices: \methodname achieves both a higher benefit-cost ratio (0.81 vs. 0.73) and much lower latency (0.741ms vs. 6.251ms, 8.4$\times$ faster). This provides strong evidence that Event Pair Set retains the structural signal needed for diversity optimization while substantially reducing computational overhead.

\begin{mybox}
\textbf{Finding 4:} \methodname achieves production-friendly efficiency with strict budget control, low per-trace latency, and high information density. Compared to an exact graph-kernel variant, EPS is both faster and yields higher benefit-cost ratio, directly validating the effectiveness of our representation and engineering optimizations.
\end{mybox}

\section{Discussion}
\label{sec:discussion}

\subsection{Applicability}

\methodname is most effective in microservice systems with multiple business entry points and relatively rich in-span events. Multiple entry points make root-span grouping more informative, while logs and span events provide the semantic signals needed to distinguish traces that are structurally similar but diagnostically different. This observation also suggests two practical considerations for deployment: root spans should reflect business entry points as clearly as possible, and trace-log correlation should be maintained consistently so that diagnostic signals remain available to the sampler.

When these conditions are weaker, \methodname's advantages become smaller but do not disappear. In systems dominated by a single entry point, such as SockShop, root-span grouping provides less leverage and the method behaves closer to a structure-driven diversity sampler. Likewise, when logs and alarms are both unavailable, \methodname falls back to its structural encoding and remains competitive with strong trace-only baselines, as shown in Figure~\ref{fig:cross_system_coverage}. Overall, \methodname benefits most from the combination of structural diversity and semantic signals, while the structural component alone remains useful.

\subsection{Threats to Validity}

\textbf{Construct Validity.} A primary threat is whether our evaluation metrics adequately capture sampling quality. We mitigate this by combining six intrinsic metrics with downstream RCA accuracy. API Coverage and Path Coverage quantify coverage at the entry-point and execution-flow levels, Trace Pattern Coverage and Shannon Entropy quantify behavioral diversity, and Proportion Anomaly and Proportion Rare quantify whether diagnostically valuable traces are preserved. 

Another concern is the choice of RCA tools. Several learning-based RCA tools performed poorly on our benchmark, likely due to domain shift from the simpler systems on which they were originally developed or evaluated. This is consistent with our previous findings on the generalization limits of learning-based RCA methods~\cite{fang_rethinking_2025}. We therefore focus the RCA study on three tools that remain stable on our dataset, while retaining the intrinsic metrics as an independent view of sampling quality.

A further threat concerns the fidelity of \methodname's semantic signals. EPS is intentionally lossy and does not preserve all timing or hierarchical information. However, EPS is used only as an in-memory proxy for sampling decisions; once a trace is selected, the complete original trace, including timestamps, topology, and raw logs, is preserved for downstream analysis. Moreover, the results in §\ref{sec:rq1} show that no evaluated system reaches 100\% Trace Pattern Coverage even at a 10\% sampling rate, indicating that the current EPS granularity is already sufficiently fine for the sampling task. Once the quota allocator has grouped traces by API endpoint, distinctions among independent events and causal relations become more important than further encoding timing or hierarchical details. In this sense, omitting those details is a deliberate trade-off for online efficiency rather than an accidental loss of information.

In addition, log parsing introduces uncertainty. We use Drain to remove dynamic fields and stabilize template IDs, since otherwise high-cardinality log contents would cause severe state-space explosion. Imperfect parsing may still over-merge or split templates. In our setting, \methodname depends more on template stability than on perfect semantic parsing, because template IDs serve as lightweight event markers rather than full log interpretations.

\textbf{Internal Validity.} Implementation bias is a potential threat in comparative evaluation. We therefore use official baseline implementations when available and otherwise follow published specifications as closely as possible. For MicroRCA, Nezha, and ShapleyIQ, we retain the default algorithmic settings from their original releases; the only changes are engineering optimizations needed to execute the evaluation pipeline at scale, especially for Nezha, without altering the underlying ranking logic. To reduce variance from stochastic factors, all metrics are computed independently for each of the 161 fault-injection cases and then averaged. Our evaluation pipeline is open-sourced to support independent verification.

\textbf{External Validity.} Our study uses open-source benchmarks, which are necessarily simpler and cleaner than many production deployments. Although the cross-system evaluation demonstrates consistent performance across diverse architectures, the advantages of \methodname are stronger in systems with multiple entry points, richer event spaces, and more complex trace patterns. Industrial systems may further exhibit noisier, delayed, or misconfigured alarms, incomplete logs, broken trace contexts, heterogeneous instrumentation quality, and more complex failure interactions. These factors may weaken the precision of alarm-driven allocation or reduce the benefit of semantic encoding. In such cases, \methodname does not fail abruptly: missing logs and alarms reduce it toward a structure-driven diversity sampler, while noisy alarms are constrained by quota caps and the diversity-preserving selector so that no single alarmed group can dominate the budget and non-alarmed groups still retain baseline coverage. We partially address these concerns by evaluating the \textit{w/o Logs w/o Alarms} variant on Dataset $\mathcal{B}$; the results in Figure~\ref{fig:cross_system_coverage} show that \methodname's structural encoding alone remains competitive with, and often superior to, strong trace-only baselines. Nevertheless, validation on industrial workloads remains necessary before making stronger claims about deployment-wide impact.

\section{Related Work}
\label{sec:related}

\textbf{Distributed Tracing Standards and Ecosystem.} Standardization efforts have driven the evolution of distributed tracing toward semantic awareness. OpenTracing~\cite{opentracing-spans} and its popular implementation Jaeger~\cite{jaeger} laid the groundwork by enabling log embedding within spans. The de facto standard, OpenTelemetry~\cite{OpenTelemetry_traces}, advances this by not only unifying trace-log correlation~\cite{opentelemetry-logging} but also actively exploring how to incorporate intra-span events into sampling strategies~\cite{opentelemetry-specificationoteps0265-event-visionmd}. \methodname provides a concrete implementation for this vision. Its lightweight EPS representation encodes both trace structure and log semantics using simple hashable bigrams, offering a practical path for event-aware sampling without complex graph-based infrastructure.

\textbf{Trace Sampling.} Mainstream online sampling has evolved from simple probabilistic selection to sophisticated strategies. Early diversity-aware samplers like \textbf{PEACH}~\cite{las-casas-weighted-2018} use hierarchical clustering. Anomaly-focused methods include \textbf{Sieve}~\cite{huang-sieve-2021}, which uses RRCF to detect structural and temporal anomalies, and \textbf{Sifter}~\cite{las-casas-sifter-2019}, which samples traces with high prediction error. \textbf{TraStrainer}~\cite{huang2024trastrainer} incorporates external metrics to prioritize traces correlated with system-level anomalies. The current state-of-the-art, \textbf{TracePicker}~\cite{xieshuaiyu-tracepicker-2025}, formulates sampling as a two-phase optimization problem for coverage and latency preservation.

A distinct line of work employs GNNs to learn powerful trace representations, including STEAM~\cite{he2023steam}, TraceCRL~\cite{TRACECRL}, and iTCRL~\cite{tian-itcrl-2024}. Notably, iTCRL is also semantically aware by modeling log events as graph nodes. However, these approaches require offline training and incur high inference overhead, rendering them unsuitable for online sampling~\cite{xieshuaiyu-tracepicker-2025,gao2025gnnsactuallyeffectivemultimodal}. \methodname is designed specifically to overcome this trade-off, providing lightweight semantic awareness in a fully online model. An indirect comparison confirms \methodname's superior balance of efficiency and effectiveness: TracePicker is shown to outperform STEAM~\cite{xieshuaiyu-tracepicker-2025}, while our results in §\ref{sec:rq1} show \methodname surpasses TracePicker on the same public dataset.

\textbf{Alternative Approaches.} Beyond sampling, other paradigms address trace volume. \textbf{TraceZip}~\cite{Tracezip} uses lossless compression via structural deduplication, a technique orthogonal and potentially complementary to sampling strategies. Others propose architectural innovations: \textbf{Mint}~\cite{Mint} captures all traces by shifting from a keep-or-discard model to one based on commonality and variability, aggregating common templates while filtering parameters. \textbf{Hindsight}~\cite{zhang2023benefit} implements retroactive sampling, analogous to a car's dash-cam, by logging all data locally and retrieving a full trace only after an anomaly is detected. These architectural concepts inspire future work in integrating \methodname's principles into collection agents.
\section{Conclusion}
\label{sec:conclusion}

This paper introduces \textbf{\methodname}, a semantically-aware trace sampler that resolves the tension between semantic richness and computational efficiency through lightweight EPS encoding, alarm-driven adaptive budget allocation, and DPP-based diversity optimization. Our evaluation demonstrates that \methodname achieves superior coverage, diversity, and anomaly capture compared to state-of-the-art baselines, enabling RCA tools to achieve higher diagnostic accuracy—in some cases, even outperforming the complete unsampled dataset. This counter-intuitive result reveals that \textbf{strategic data curation can enhance signal quality beyond mere data quantity}. By bridging inter-span structure with intra-span semantics while maintaining computational tractability, \methodname transforms trace sampling from passive volume reduction into active signal enhancement for downstream diagnostics. We provide an open-source benchmark dataset (161 fault cases, 1.4M+ traces, 125K events) to support future AIOps research, demonstrating that careful design enables sampling less data while providing better insights.

\section{Data Availability}
\label{sec:data-availability}
The \methodname implementation is publicly available at~\cite{operationspaigleaner_2026}, and the benchmark dataset introduced in this paper is publicly available at~\cite{yang_2025_19637628}.

\bibliographystyle{ACM-Reference-Format}
\bibliography{ref}

\end{document}